\documentclass[twocolumn]{aastex631}

\begin{document}

\title{Thermal Structure and Millimeter Emission of Protoplanetary Disk with embedded protoplanets from  radiative transfer modeling}

\correspondingauthor{Felipe Alarc\'on}
\email{falarcon@umich.edu}

\author[0000-0002-2692-7862]{Felipe Alarc\'on }
\affiliation{Department of Astronomy, University of Michigan,
323 West Hall, 1085 S University, Ave.,
Ann Arbor, MI 48109, USA}

\author[0000-0003-4179-6394]{Edwin A. Bergin}
\affiliation{Department of Astronomy, University of Michigan,
323 West Hall, 1085 S University, Ave.,
Ann Arbor, MI 48109, USA}



\begin{abstract}

The discovery of protoplanets and circumplanetary disks provides a unique opportunity to characterize planet formation through observations. Massive protoplanets shape the physical and chemical structure of their host circumstellar disk by accretion, localized emission, and disk depletion. In this work, we study the thermal changes induced within the disk by protoplanet accretion and synthetic predictions through hydrodynamical simulations with post-processed radiative transfer with an emphasis on radio millimeter emission. We explored distinct growth conditions and varied both planetary accretion rates and the local dust-to-gas mass ratios for a protoplanet at 1200 K. The radiative transfer models show that beyond the effect of disk gaps, in most cases, the CPD and the planet's emission locally increase the disk temperature. Moreover, depending on the local dust-to-gas depletion and accretion rate, the CPD presence may have detectable signatures in millimeter emission. It also has the power to generate azimuthal asymmetries important for continuum subtraction. Thus, if other means of detection of protoplanets are proven, the lack of corresponding evidence at other wavelengths can set limits on their growth timescales through a combined analysis of the local dust-to-gas ratio and the accretion rate.
\end{abstract}



\section{Introduction} \label{sec:intro}

The observational detection of protoplanets through different means has become a reality in the present decade. A handful of protoplanets have been confirmed or nominated through direct imaging methods, including PDS 70 b and c \citep{Keppler..2018, Haffert..et..al..2019}, AB Aur b \citep{Currie..2022} and HD 169142 b \citep{Gratton..2019, Hammond..2023}. Additionally, a larger number of protoplanet candidates have been proposed through velocity ''kinks''  and footprints in disk gas kinematics \citep{Pinte..2019, Pinte..2020} and more recently via detections of circumplanetary disks (CPDs) in gas or dust emission \citep{Benisty..2021,MAPS..XXI..Bae}. While existing facilities (e.g. Atacama Large Millimeter Array/ALMA, the James Webb Space Telescope/JWST, and ground-based optical/infrared telescopes) will continue to push this field, the upcoming large ground-based optical observatories, such as the European ELT, the Thirty Meter Telescope, and the Giant Magellan Telescope will further revolutionize our knowledge of planet formation. This increased observational capacity can potentially help to achieve multiple wavelength characterization of young planet-forming systems. This means we will be able to observe and analyze (proto)planets and their natal disks across a wide range of wavelengths, ranging from optical light to infrared. 

 During their formation, giant planets change the physical and chemical conditions in their surroundings by carving gaps in the gas and dust which leads to radial differences in the temperature and radiation fields \citep{ Facchini..2018, van..der..Marel..2018, Alarcon..2022, Broome..2023}. They cause chemical exchange between layers through distinct gas and dust meridional flows \citep{Szulagyi..et..al..2014,Dong..2019, Szulagyi..et..al..22..flows} and locally elevate the temperature in the surrounding disk potentially inducing changes in the chemical composition \citep{Cleeves..2015}.  Much of the extant work in the references above is based on ALMA images of gas and dust in disk combined with theoretical models. However, additional observational constraints are becoming available.
  As an example, the detection of UV continuum towards PDS 70 b supplies information about its accretion rate and possibly about its Circumplanetary Disk or CPD \citep{Yifan..2022}. From a chemical point of view, determining the effective accretion rate is important to determine the UV and H$\alpha$ luminosity, which sets up the thermal and photochemical equilibrium of the planet-feeding gas, where atomic tracers may be dominant \citep{Alarcon..2022}.


 In this investigation, we explore the thermal effects of a growing planet and the expected emission at millimeter wavelengths by post-processing models with different accretion rates and local levels of dust depletions inside, and outside, the millimeter/pebble disk.  We then use these predictions to study the effective protoplanet footprint within the current capabilities of ALMA. 

This paper is organized as follows: we present the methodology applied in the work to study the radiative transfer effect of a planetary source in a circumstellar disk in Section \ref{sec:methods}. In Section \ref{sec:results} we present the main results of our simulations exploring different parameters relevant in planet-forming settings. We discuss the main findings of our work in Section \ref{sec:disc} and we summarize the present paper in Section \ref{sec:summary}.

 \section{Methods} \label{sec:methods}

We explore the thermal differences associated with the emission of a protoplanet within a protoplanetary disk. We first run 3D hydrodynamical simulations with \texttt{FARGO3D} \citep{FARGO3D}, and then post-process these simulations with \texttt{RADMC-3D}\citep{RADMC-3D} by including the blackbody emission of a protoplanet, accretion shocks, and the spectrum of a circumplanetary disk. We then produced synthetic continuum emission images of the hosting disk at $\lambda$=1.3 mm observable by ALMA at typical angular resolutions of dust continuum emission, $\sim$ 3 au \citep{Andrews..20}.

\subsection{Hydrodynamical Simulation Setup}

We ran two 3D hydrodynamical simulations with \texttt{FARGO3D} \citep{FARGO3D}, which is built expanding the  \texttt{FARGO} algorithm \citep{Masset2000}, particularly efficient for advection flows in Keplerian disks. Using a dedicated hydrodynamics code provides a more realistic physical setup for the radiative transfer models. One of them replicated the best-fit 2D model of the HD 163296 dust continuum emission from the DSHARP survey \citep{DSHARP..I, DSHARP..VII}. We also embedded a planet beyond the edge of the observable millimeter disk for the other run. We decided to replicate the HD 163296 disk as our starting point as it presents a well-known and thoroughly characterized protoplanetary disk, particularly at high resolution in millimeter wavelengths as part of both the DSHARP and MAPS surveys \citep{DSHARP..I, MAPS..I..Oberg}. The HD 163296 disk presents  a rich substructure in dust emission with crescents, dark dust gaps, and bright emission rings \citep{DSHARP..IX}. Given this structure, it has been proposed in the literature that the HD 163296 disk hosts several protoplanets that are predicted via simulations that match the structures seen in dust continuum emission  \citep{DSHARP..VII, DSHARP..IX, Garrido-Deoutelmoser..2023}.
Additional evidence for protoplanets is found in the
gas kinematics acting as planetary footprints \citep{Pinte..2019, Pinte..2020, Teague....et..al..2019, Izquierdo..2022, Izquierdo..2023}; and from the emission of atomic tracers \citep{Alarcon..2022}. In particular for our hydrodynamical run, we located three planets of 0.7, 2.18 and 0.14 M$_{\rm Jup}$ at 10, 48 and 86 au respectively, although we focus on the effect of the planet at 48 au, which is   the most massive inside the millimeter disk. 

\begin{figure}[h!]
    \centering
    \includegraphics[width=0.99\linewidth]{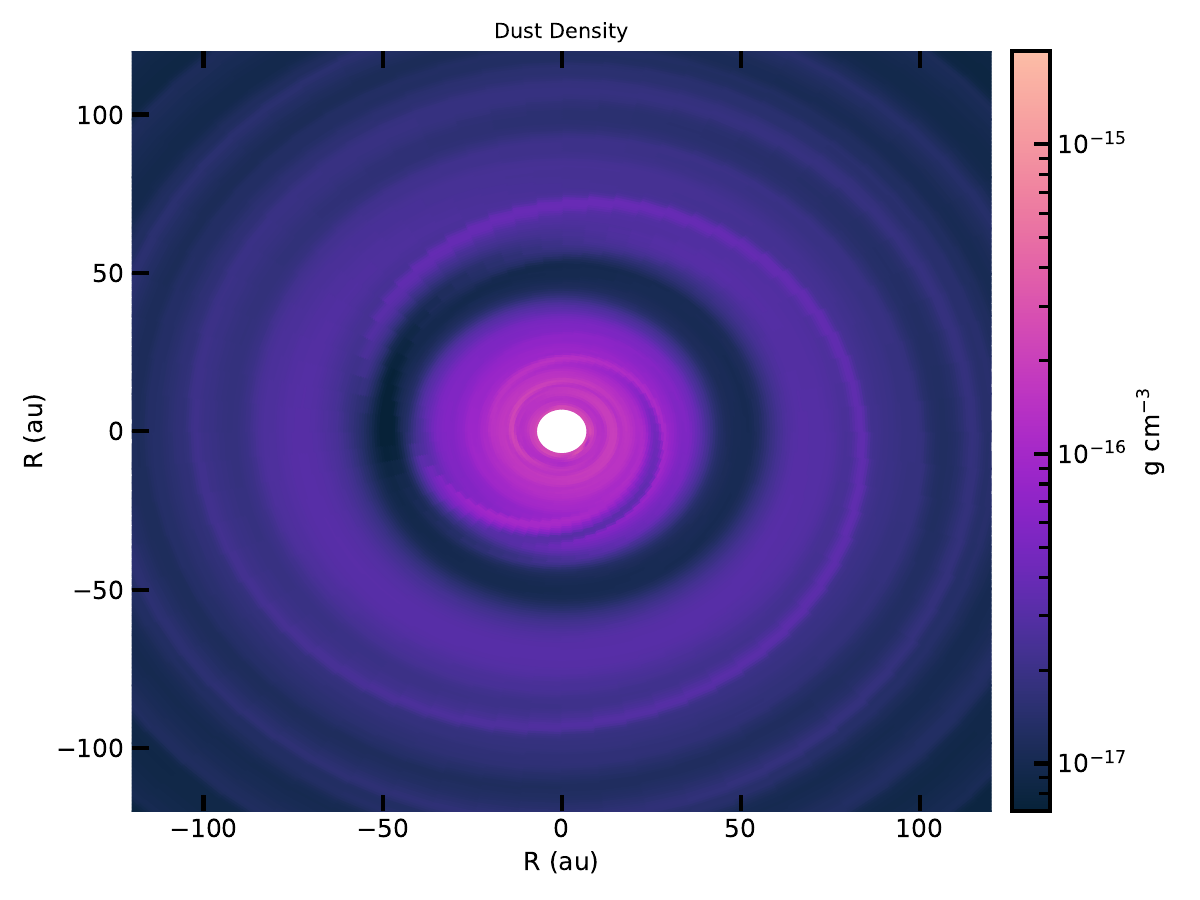} \\
    \includegraphics[width=0.99\linewidth]{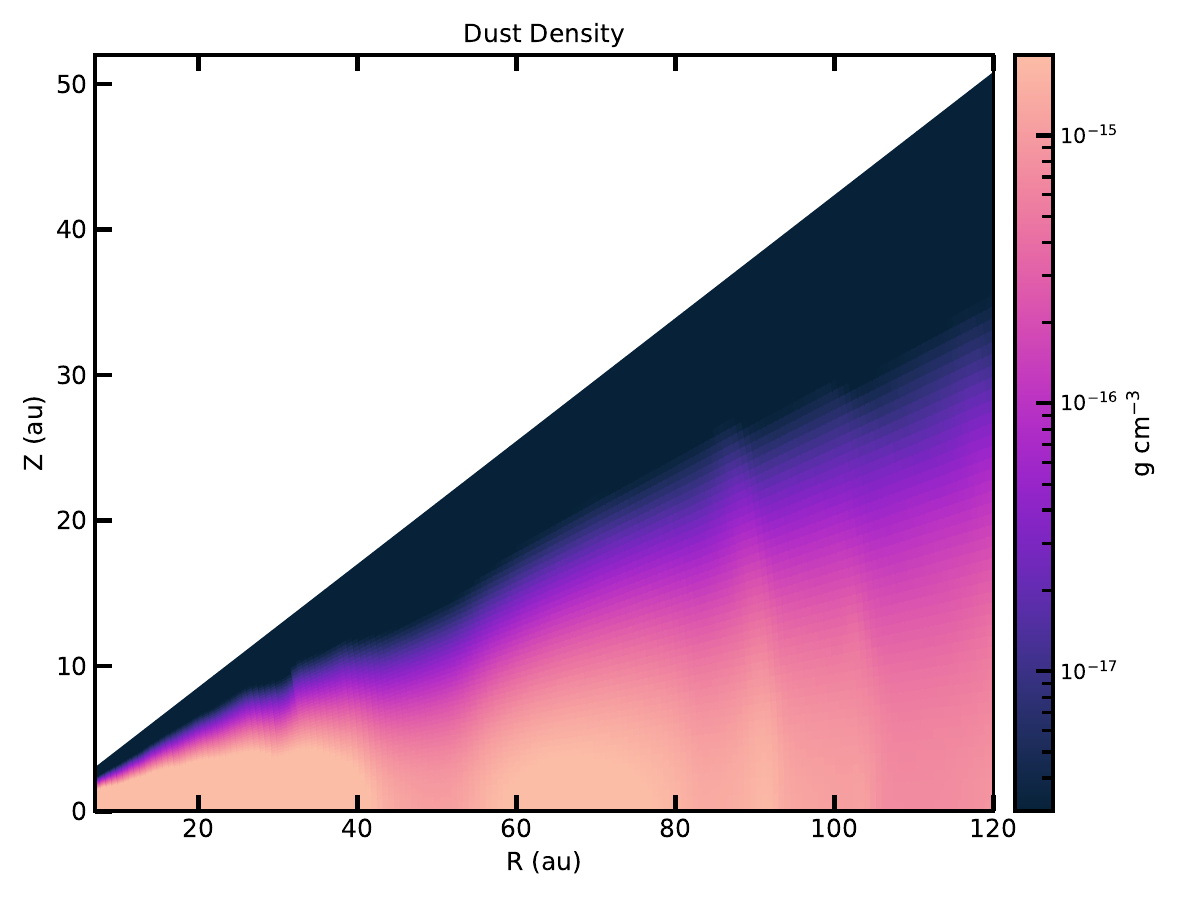}
    \caption{\textbf{Top:} Density field in the midplane of the disk after 500 orbits of the HD 163296 model using the \texttt{FARGO3D} code \citep{FARGO3D}. \textbf{Bottom:} Meridional cut of the density field along the azimuth of the planet at 48 au showing the gap and density structure of the disk in the radial and vertical direction. }
    \label{fig:hydro}
\end{figure}

 In our second modeling setup, we analyze the effects of a circumstellar disk with a massive planet located beyond the edge of the millimeter dust disk, i.e., where dust emission falls below a given sensitivity limit. In this model, we embedded a 2 M$_{\rm Jup}$ planet at 150 au beyond the observable millimeter disk. It is expected that the effects, in this case, will be more noticeable and easier to decouple from the stellar-disk interactions as the disk becomes colder in the outer regions. A mass tapering for planetary growth in both models was included over the first 100 orbits.

Our models assume a constant $\alpha$-viscosity prescription  \citep{Sakura-Sunyaev} across the disk with $\alpha=10^{-3}$. The 3D grid for the two hydrodynamical simulations have (256,512,64) cells in spherical coordinates ($r$,$\phi$,$\theta$). The colatitude $\phi$ spans from 1.17 to $\frac{\pi}{2}$ assuming mirror symmetry with respect to the midplane, while the azimuthal coordinate covers the whole range from 0 to 2$\pi$. The radial coordinate ranged from 7 to 200 au for the HD 163296 disk model and from 20 to 400 au for the single-planet model with a logarithmic spacing. The radial velocity has antisymmetric boundary conditions in the radial edges and symmetric in both ends of the colatitude range, while the colatitude velocity is symmetric in radius and antisymmetric in the colatitude boundaries. The hydrodynamical simulation ran for 500 orbits of the most massive planet at 48 au in the HD 163296 model and for the same number of orbits for the model with one planet at 150 au. An aspect ratio of 0.08 with a flaring index $\psi=$0.08 was considered in the initial density distribution of the disk. The planets were input in fixed circular orbits using the planet-to-star mass ratio, $q$, assuming a stellar mass of 2.1 M$_{\odot}$. We show the dust density fields of our HD 163296 3D hydrodynamical simulation in Figure \ref{fig:hydro}. In the simulation, we see the density depletion in the gaps as expected in the location of the massive planets along with the planet-driven spirals.

\subsection{Radiative Transfer Setup}

We studied the effects of the emission of a protoplanet and respective CPD by post-processing the output from the hydrodynamical simulations of \texttt{FARGO3D} with \texttt{RADMC-3D} \citep{FARGO3D, RADMC-3D}. 
Within the heating components in the modeling, we took into account the viscous heating associated with the stellar accretion rate, assuming a stellar accretion rate of $\dot{M}= 0.69 \times 10^{-7}  M_{\odot} \ $yr$^{-1}$ \citep{Donehew..2011}. The viscous heating was modeled by adding concentric rings following the standard temperature distribution of an accretion disk \citep{Calvet..1994}, 

\begin{equation}\label{visc}
        q_{\rm visc} = \frac{3GM_P\dot{M}_P}{8\pi R^3}\Big[1 - \big(\frac{R_P}{R}\big)^{1/2} \Big].
\end{equation} 
        
\begin{figure}[!h]
    \centering
    \includegraphics[width=1\linewidth]{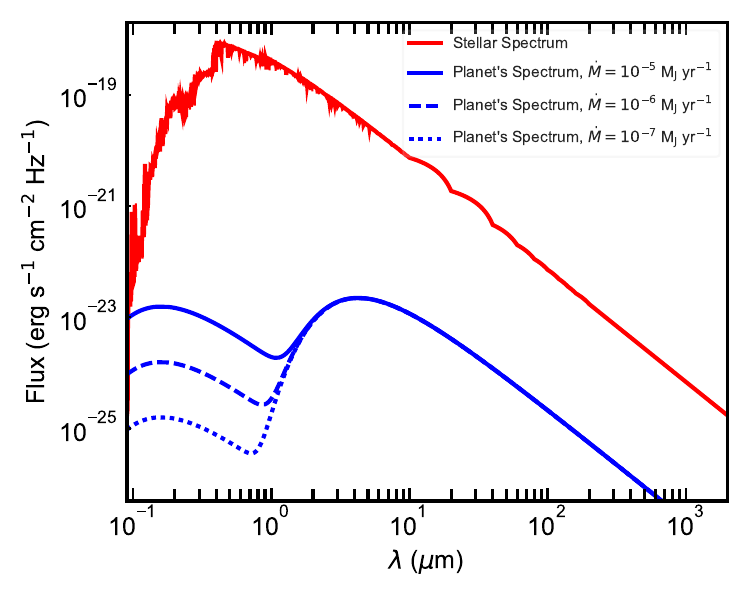}
    \caption{Input spectra of the model for different planetary accretion rates, which includes the circumplanetary disk and the accretion shocks. The stellar spectrum of the HD 163296 star is also shown for reference and comparison with the protoplanetary spectra.}
    \label{fig:spec}
\end{figure}

\noindent We used 10$^8$ photons to calculate the thermal equilibrium. The stellar spectrum is the one corresponding to the HD 163296 star given that we set up the hydrodynamical simulations based on that disk. The input spectra for the star and the protoplanetar are shown in Figure \ref{fig:spec}.

We modeled the dust opacity in the disk using two dust populations for the radiative transfer modeling. The first population corresponds to large grains holding 99\% of the disk dust mass, and the other population represents only small grains with a 1\% of the disk dust mass. Both populations have the same composition, which corresponds to a water ice mass fraction of 20\%, 40\% in organics, 33\% astrosilicates, and $\sim$ 8.8\% in troilites. The opacity values were taken from \cite{Warren..Brandt..2008} for water ice, \cite{Henning..Stognienko..1996} for organics and troilite, and \cite{Draine..2003} for the astrosilicates.

We set a constant  dust-to-gas mass ratio across the disk of 0.01 \citep{Sandstrom..2013} with the exception of the Hill sphere around the planet at 48 au, which is a free parameter in our modeling that will be detailed below. \cite{Zhu..2012} demonstrate that planet-induced gaps filter the drifting of large grains; additionally, \cite{Zhu..2018} show that the drifting timescale of millimeter and sub-millimeter dust grains in massive CPDs can be as short as $t_{\rm drift}\sim$ 1000 yr.  Thus, we explored different scenarios varying the local depletion of large grains, which is reflected in the local dust-to-gas-mass ratio, assuming values of 10$^{-2}$,10$^{-3}$ and 10$^{-4}$. As part of the radiative transfer simulation with  \texttt{RADMC3D}, anisotropic scattering with the Henvey-Greenstein phase function was considered \citep{Hen..Green}. The dust size distribution for the two populations follows the \cite{Mathis..1977} power law proportional to the dust grain size, i.e., 

\begin{equation}
    n_0(a) \propto a^{-p}da,
\end{equation}

\noindent with $a$ the grain size, and $p$ the power law index, which we set to 3.5. The first dust population ranges from $a_{\rm min}$ = 0.05 $\mu$m to $a_{\rm max}$ = 2.5 mm, while the second dust population spans sizes between $a_{\rm min}$ = 0.05 $\mu$m to $a_{\rm max}$ = 5  $\mu$m.

\subsubsection{Protoplanetary Spectra}

We added an additional point source representing the emission from the protoplanet and its associated CPD. The spectrum of the planet is a combination of three components:

\begin{enumerate}
    \item Thermal emission coming from the protoplanet itself which is modeled as a blackbody emitting with a temperature of T$_p$ = 1200 K.
    \item The emission coming from the accretion of the CPD using the viscous heating from Equation \ref{visc} inside the Hill's sphere of the planet. For this case, viscous heating will be dependent on the mass of the planet,  $M_P$, the planetary accretion rate, $\dot{M}_P$, treated as a free parameter, $R_P$, the radius of the planet, and the separation between the star and the planet, $R$.
    \item The shock emission is estimated following the prescription of \cite{Calvet..Gullbring..98} and \cite{Zhu..2015}. The temperature of the shocked gas, $T_{\rm sh}$, is:

    \begin{equation}
        T_{\rm sh} = \frac{3}{16}\frac{\mu m_H}{k}v_s^2 = 3.44 \times 10^6 \ \mathrm{K} \Big(\frac{M_P}{M_{\odot}} \Big)\Big(\frac{R_P}{R_{\odot}} \Big)^{-1},
    \end{equation}

    \noindent where $m_H$, $\mu$, $k$ are the hydrogen mass, the mean molecular weight, and the Boltzmann constant respectively; $v_s$ is the free-fall velocity at the planetary surface:

    \begin{equation}
        v_s = \Big(\frac{2GM}{R_P} \Big)^{1/2}\Big(1 - \frac{R_P}{R_i} \Big)^{1/2},
    \end{equation}

\noindent with $G$ the gravitational constant, $R_P$ the planetary radius, and $R_i$ the inner radius of the accretion disk. 

\end{enumerate}

 We added an extra emitting point source in the radiative transfer modeling to explore the effects of a massive protoplanet thermal emission, the CPD, and shocks emission. The additional point source corresponds to the most massive planet at the same location as the hydrodynamical simulation, i.e., at 48 au for the HD 163296 model and at 150 au for the single planet one. The SED of the planets is dependent on multiple factors such as the accretion rate, $\dot{M}$, and the assumed effective shock area over the surface of the planet. Hence, we explored three different values for the accretion rate: $\dot{M}$ =10$^{-7}$, 10$^{-6}$, and 10$^{-5}$ M$_{\odot}$ yr$^{-1}$.

\begin{figure*}
    \centering
    \includegraphics[width=1\linewidth]{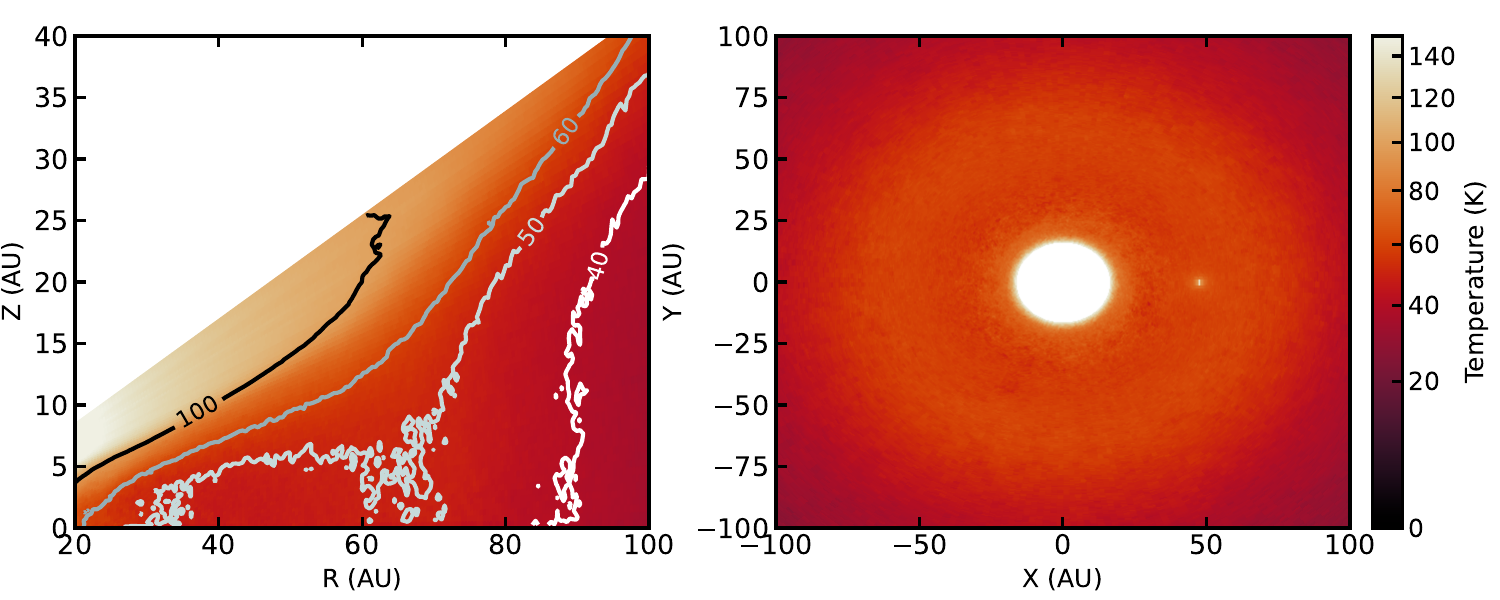}
    \caption{Dust temperature in the disk for a with a planet having an accretion rate of $\dot{M}=10^{-5}\ M_{\mathrm{Jup}}$ yr$^{-1}$ showing the local increase of the temperature and its extension in the three directions where the protoplanet is located. \textbf{Left:} Dust temperature in a meridional cut along the azimuthal location of the protoplanet. \textbf{Right:} Dust temperature in the midplane of the disk.}
    \label{fig:T_1200}
\end{figure*}

To account for the drift of large dust \citep{Zhu..2015} we also change the local dust-to-gas ratio inside the typical size of a CPD, which is around one-third of the planet's sphere of influence, i.e., the Hill sphere \citep{Ayliffe..Bate..2009, Szulagyi..et..al..2014, Perez..2015}. The Hill radius, $r_H$, in the circular orbits of the embedded planets in our simulations is characterized by:

\begin{equation}
    r_H = a\sqrt[3]{\frac{M_P}{3M_*}},
\end{equation}

\noindent were $a$ is the orbital radius of the planet, $M_P$ the planetary mass, and $M_*$ the stellar mass, which in our case matches the mass of HD 163296, i.e., $M_*$ = 2.1 $M_{\odot}$ \citep{MAPS..I..Oberg}. Given our assumption of a 2 M$_{\rm Jup}$ mass planet at 48 and 150 au, the Hill sphere has a size of  3.27 au and 10.2 au for each model respectively. We used different values for the dust-to-gas ratio inside the CPD, adopting three different values; an ISM value of \textit{d2g}=10$^{-2}$, a moderate depletion value  \textit{d2g}=10$^{-3}$, and a severely dust-depleted case with \textit{d2g}=10$^{-4}$.

\subsubsection{Thermal Equilibrium and Synthetic Images}

We post-processed the hydrodynamical simulations by creating synthetic images using \texttt{RADMC-3D} \citep{RADMC-3D} to study the observability of the distinct accretion and CPD setup in radio wavelengths. We created 1.3 millimeter dust continuum images ALMA assuming a disk PA of 43.4 degrees and an inclination of 46.6 degrees. These angles follow the respective fitted values from high-resolution gas kinematics observations of the HD 163296 disk \citep{MAPS..XVIII}. We used the \texttt{SIMIO} package \footnote{https://github.com/nicokurtovic/SIMIO} to reproduce the observing conditions of the HD 163296 disk, matching the antenna configuration, the coverage of the \textit{uv} space and the time integration.


It is expected that due to the lower optical depth, the temperature inside dust-depleted gaps is hotter than the one of a smooth disk \citep{Facchini..2018, van..der..Marel..2018, Alarcon..et..al..2020}, which has been observationally confirmed to be true for HD 163296 \citep{MAPS..XVII}. Thus, to isolate and avoid mixing the effect of the gap itself, we compare the thermal fields with the hydrodynamic simulation with the planet embedded turning on and off the CPD and planetary emission. We apply the same criterion for the synthetic ALMA continuum images at 1.3 mm where those effects have already been considered.

The resulting synthetic ALMA images were then subtracted and compared with a post-processed hydrodynamic simulation with planet-driven dust-depleted gaps but without the presence of a secondary emitting source.

\section{Results}\label{sec:results}

\subsection{Temperature Structure}

We recover the expected thermal changes of previous studies by post-processing hydrodynamical simulations with radiative transfer, i.e.,  the temperature increases around the planet's location by tens of Kelvin degrees \citep{Montesinos..2015, Cleeves..2015, Portilla_Revelo..2022, Oberg..Kamp..2022}. Our work expands on those studies by splitting the additional energy source between the blackbody emission from the planet, accretion shocks, and the CPD itself (see Figure \ref{fig:spec}).

In Figure \ref{fig:T_1200} we illustrate the thermal fields in a meridional cut ($r$,$z$)  at the planet's location and the ($x$,$y$) thermal field in the midplane from the models with the planet's temperature $T_P$= 1200 K and a high planetary accretion rate, $\dot{M}=10^{-5}$ M$_{\rm Jup}$ yr$^{-1}$. The planet point source emission effectively changes the physical temperature in the surroundings in physical scales up to $\sim$ 5 au at 48 au, i.e., at least one disk scale height. The extension of this thermal effect is comparable to the size of the Hill's sphere.

\begin{figure*}[h!]
    \centering
    \includegraphics[width=1.0\linewidth]{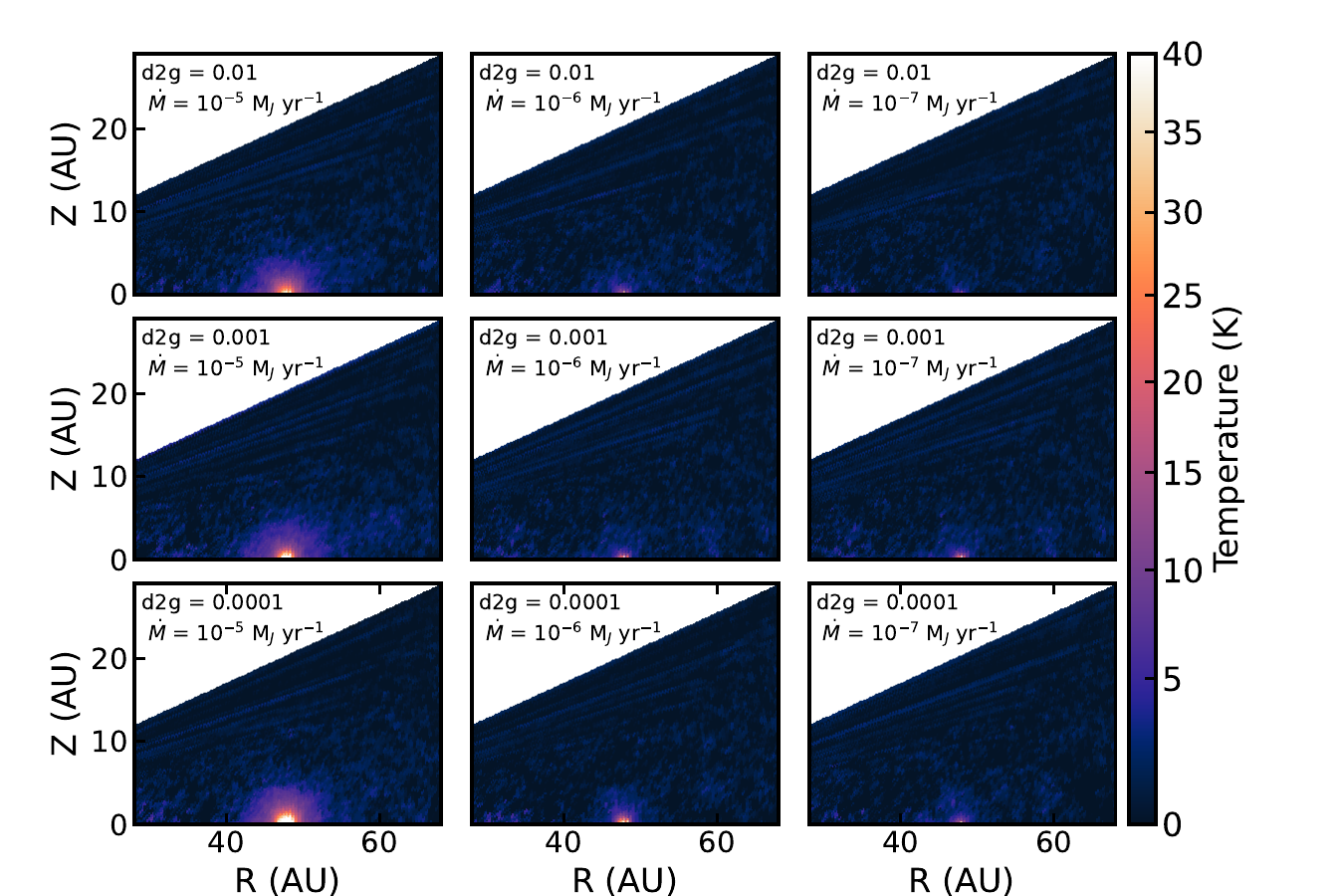}
    \caption{Dust temperature changes in the disk caused by the additional input spectra for the different conditions of the protoplanet and protoplanetary accretion rates along meridional cuts at the location of the planet. The differences show that for accretion rates $\dot{M}>10^{-7}  M_{\mathrm{Jup}}$ yr$^{-1}$ the thermal increase is noticeable while for accretion rates $\dot{M}\leq 10^{-7}\ M_{\mathrm{Jup}}$ yr$^{-1}$ is probable that the effect is milder and not significant even taking into account the blackbody emission of the protoplanet at 1200 K.}
    \label{fig:dT_1200}
\end{figure*}

\begin{figure*}[h!]
    \centering
    \includegraphics[width=1.0\linewidth]{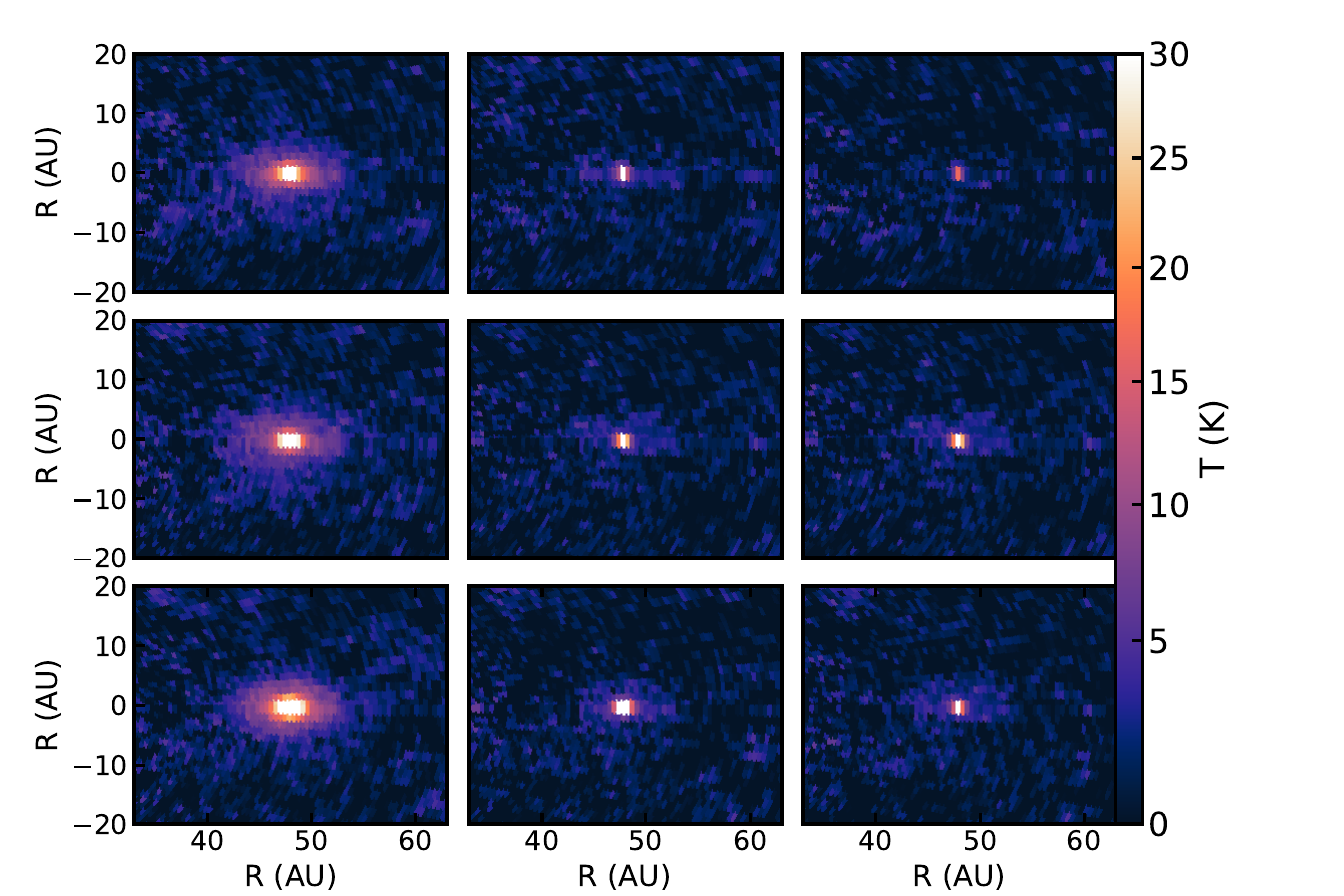}
    \caption{Dust temperature changes in the disk caused by the additional input spectra for the different conditions of the protoplanet and protoplanetary accretion rates in the midplane of the disk. The differences show that for accretion rates $\dot{M}>10^{-7}$ M$_{\mathrm{Jup}}$ yr$^{-1}$ the thermal increase is noticeable while for accretion rates $\dot{M}\leq 10^{-7}\ M_{\mathrm{Jup}}$ yr$^{-1}$ is probable that the effect is milder and not significant even taking into account the blackbody emission of the protoplanet at 1200 K.}
    \label{fig:dT_x_y_1200}
\end{figure*}

We show a comparison of the models with the extra emitting sources at 48 au for the different setups in Figures \ref{fig:dT_1200} and \ref{fig:dT_x_y_1200}. From the thermal distributions, we observe that in all cases there is a noticeable effect in azimuth and height for accretion rates  $\dot{M}>10^{-7}$ M$_{\odot}$ yr$^{-1}$, indistinct of the local large dust depletion. The expected trend in temperature with the planetary accretion rate is confirmed. As the planetary accretion rate increases, the CPD and shock emission does as well, producing an extra local heating effect. For $\dot{M}=10^{-7}$ M$_{\odot}$ yr$^{-1}$, the effects are almost negligible with very small changes in the vicinity of the planet. It is noticeable that very high accretion rates $\dot{M}>10^{-6} $M$_{\odot}$ yr$^{-1}$ produce strong thermal signatures, i.e., the temperature increases by at least 20 K. Such thermal increases are significant to change the chemical balance in the gaseous and solid composition.

When looking at the trend of different local dust-to-gas ratios, the effect is slightly more significant for the models with severe dust depletion values. As dust depletion increases, the surrounding material around the planet becomes more transparent increasing the mean free path of emitted photons from the planet or the CPD making the thermal gradient flatter, although the overall extension of the thermal changes does not change drastically. However, the thermal changes within the Hill sphere are more uniform for the lowest dust-to-gas ratios. It is worth mentioning that the same increased transparency will cause an enhanced local UV flux with a considerable extension. This topic and a more careful UV radiative transfer will be the subject of a subsequent paper.

   The thermal increases in the disk may have a significant effect that can create a localized sublimation effect, further creating azimuthal asymmetries in the disk. However,  such thermal changes are not critical when considered together with the temperature  gains due to the gap itself \citep{van..der..Marel..2018, Alarcon..et..al..2020}. Figures \ref{fig:dT_1200} and \ref{fig:dT_x_y_1200} also show that the extension of the thermal differences is dependent on the parameters of the CPD, and there is an overall thermal increase in the disk due to the protoplanetary emission; however, this net thermal increase should not be significant enough as it is of the order of 1-2 K in regions far from the planet. When the temperature field is compared between azimuthally opposed extremes of the disk, the difference is only noticed as a bright hot spot surrounding the planet. Therefore, the overall 1-2 K increase across the disks within 20-70 au from the star is coming from the blackbody emission of the protoplanet.  This effect is not really detectable with current state-of-the-art observatories, but it points to an overall heating of the disk just produced by the localized blackbody emission from the planet.

\begin{figure*}[h!]
    \centering
    \includegraphics[width=1.0\linewidth]{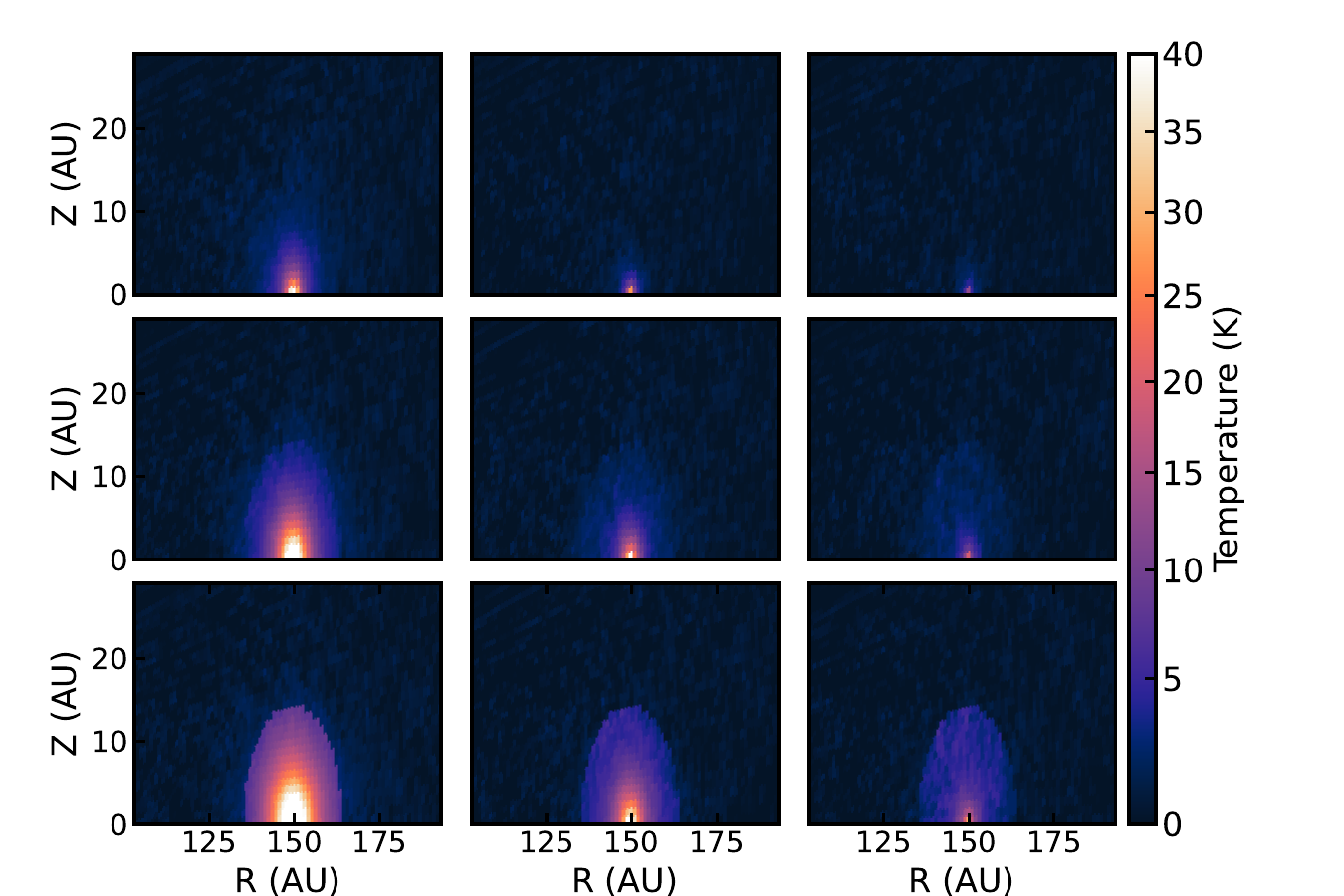}
    \caption{Meridional cut at the azimuth where the planet is located. The trends shift from the embedded planet compared to when is located at 48 au is caused mainly by changes in the dust optical depth.}
    \label{fig:dT_1200_150}
\end{figure*}

\begin{figure*}[h!]
    \centering
    \includegraphics[width=1.0\linewidth]{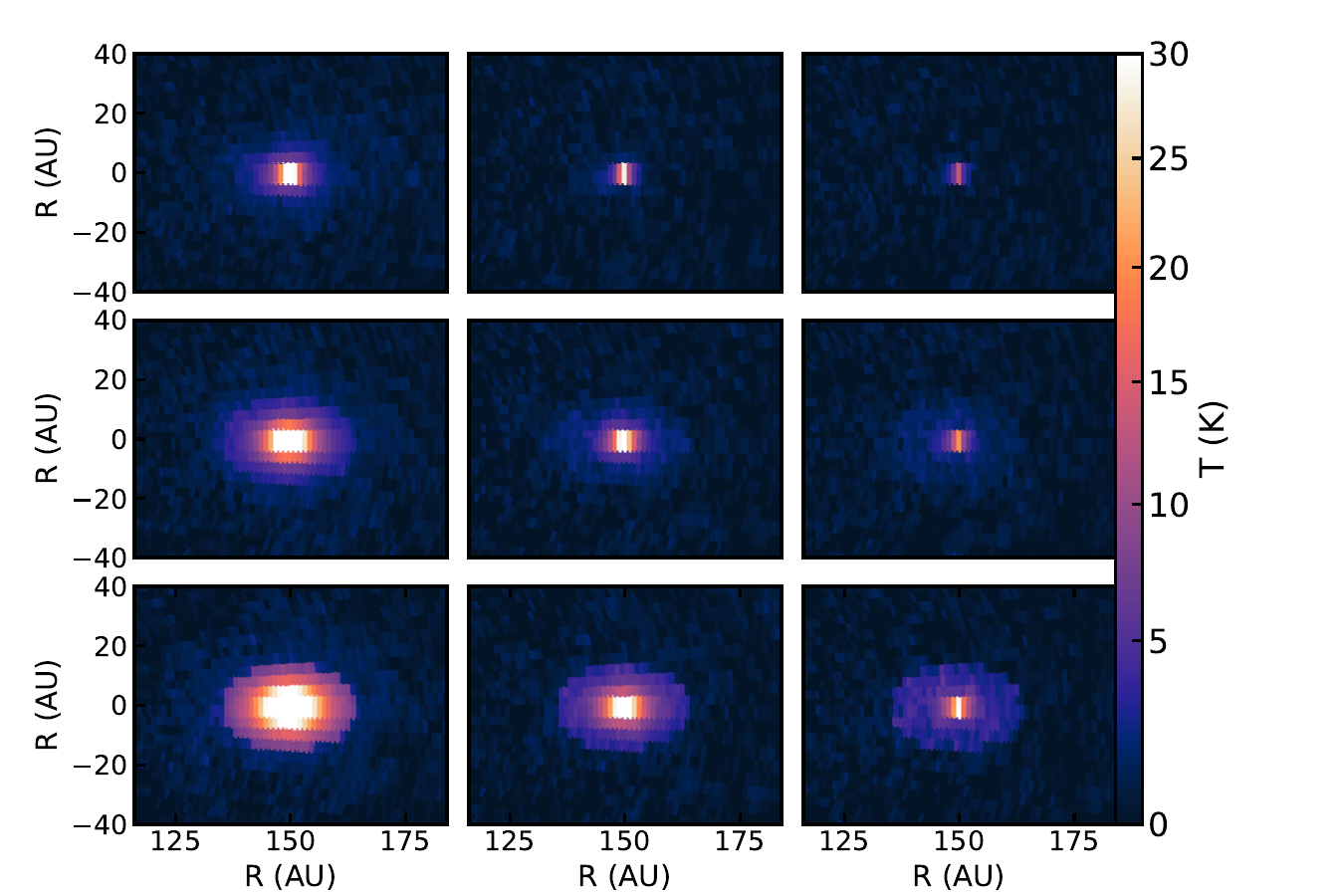}
    \caption{Midplane temperature changes with the simulation of a single planet embedded at 150 au. Dust temperature changes when the planet is embedded beyond the edge of the pebble disk  at 150 au.}
    \label{fig:dT_1200_150_2}
\end{figure*}

 In Figures \ref{fig:dT_1200_150} and \ref{fig:dT_1200_150_2}, we show the same thermal differences but compare the single-planet model at 150 au instead of the HD 163296 disk. We observe that many of the trends observed in the simulation with the massive planet inside the pebble disk persist.  One of those trends is that by boosting the planetary accretion rate, the local temperature rises as the overall energy input from the shocks and CPD increases. However, there are also important distinctions related to the extension of these changes and their absolute values.

 One of the main differences is the trend with the local dust-to-gas ratio. With a similar planetary accretion rate, local dust depletion has the potential to produce drastic changes in the disk temperature around the planet that is found beyond the edge of the dust disk. Such changes can be extended to more than 10 au with temperature increases over 25 K. Thus, these changes are within achievable spatial scales resolvable with ALMA. Additionally, in the presence of a massive gaseous CPD, or detectable line emission, molecular tracers could probe changes in the local line ratios or in the abundance of volatile species. 

 When the planet is beyond the pebble disk, the overall heating of 1-2 K degrees over the surrounding material disappears. The probable reason behind this phenomenon is the lack of scattering processes of the blackbody emission due to a much lower dust density, i.e., the mean free path of dust scattering is large enough for the thermally emitted photons to escape the disk.

\subsection{Millimeter synthetic emission}

\begin{figure*}[h]
    \centering
    \includegraphics[width=1.0\linewidth]{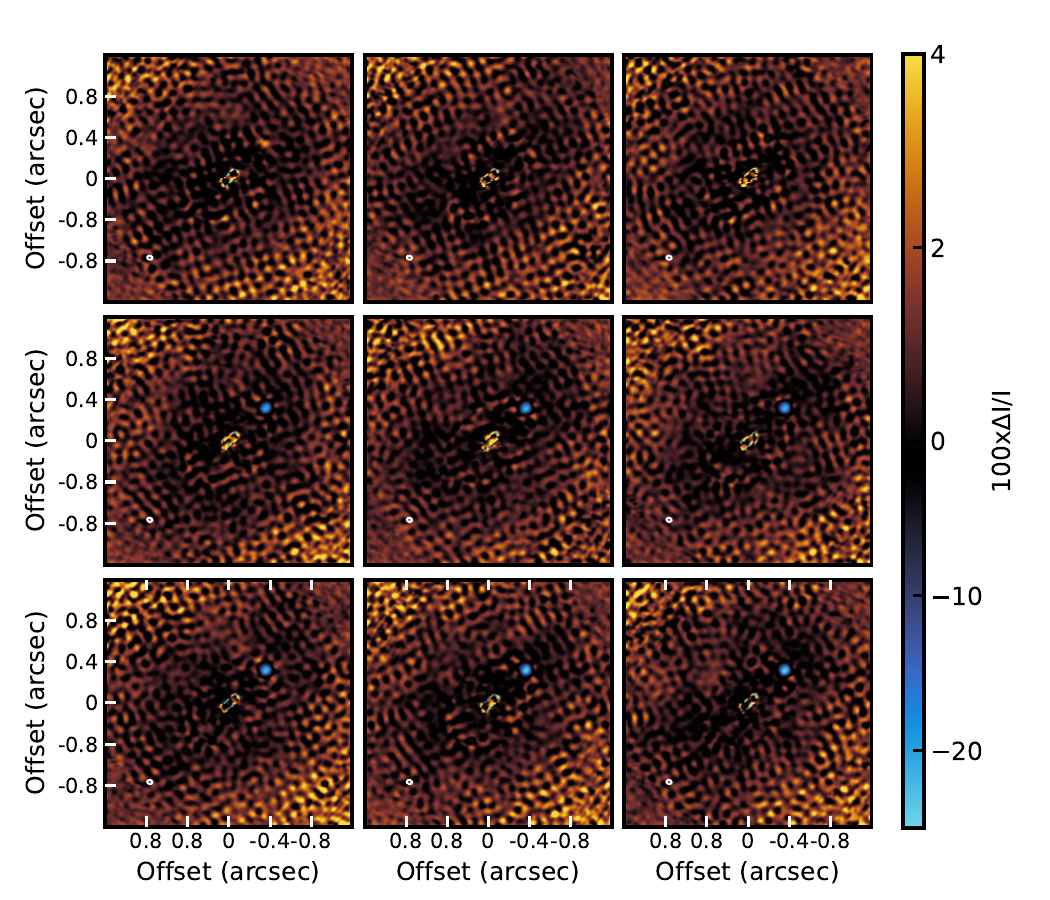}
    \caption{Comparison of synthetic 1.3-millimeter dust continuum images using \texttt{SIMIO} between the different models and a standard model without protoplanetary emission for  $\sim$1 hour integration time. Except one case, there is no strong additional signature of the CPD. Furthermore, when the CPD is severely depleted of large grains, the dust millimeter emission is lower than the surroundings. From the low accretion rate models we also infer that the 1200 K blackbody protoplanetary spectrum does not produce a significant change in synthetic ALMA images}
    \label{fig:ALMA}
\end{figure*}
\begin{figure}[h!]
    \centering
    \includegraphics[width=0.99\linewidth]{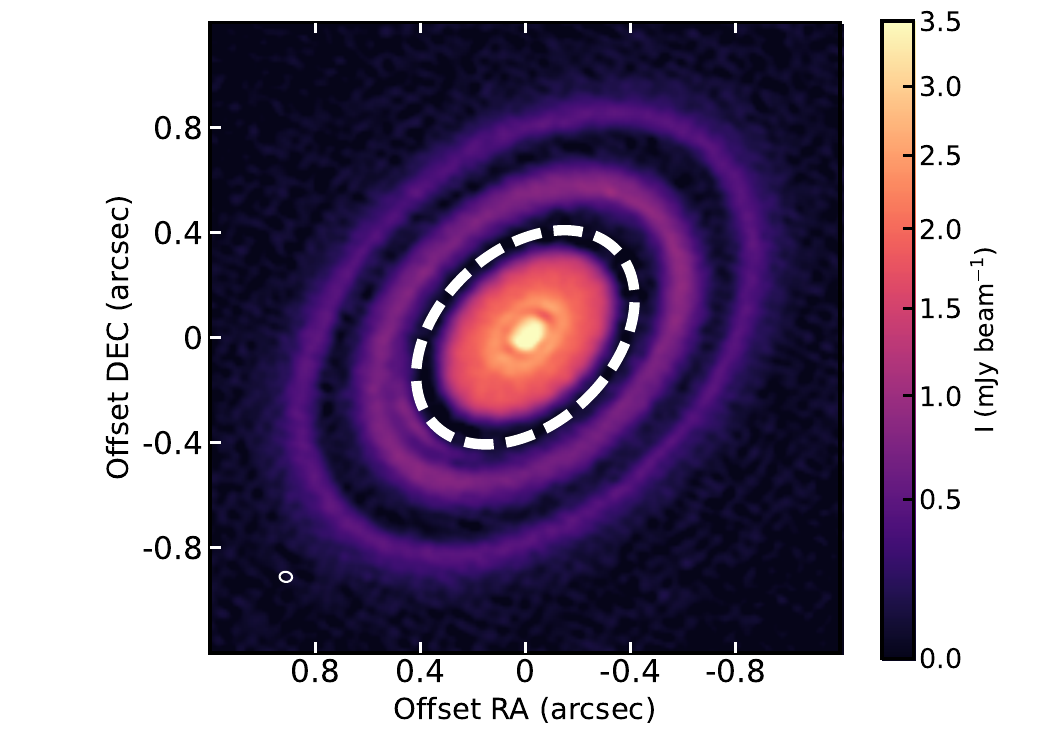}
    \includegraphics[width=0.99\linewidth]{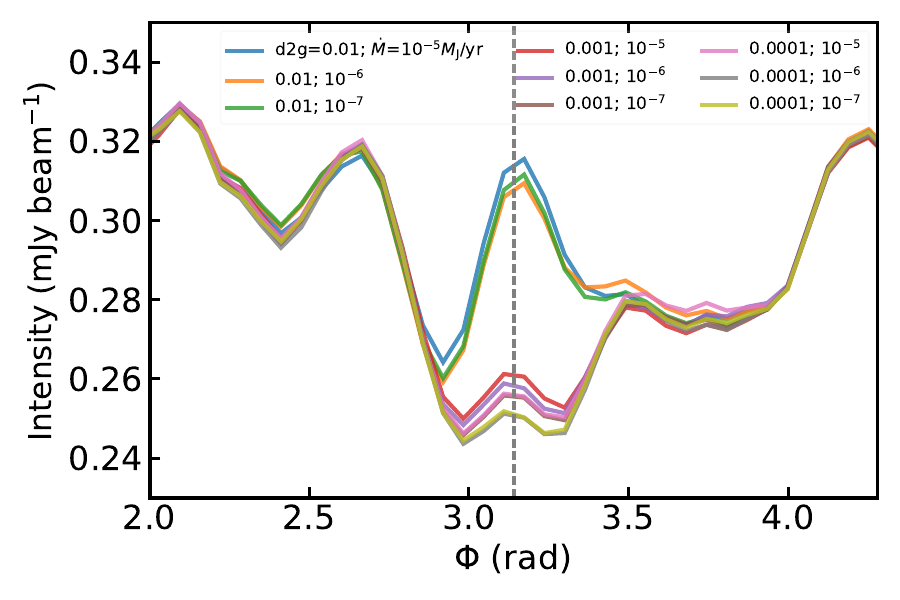}
    \caption{\textbf{Top:} 1.3 millimeter continuum emission image from the ALMA DSHARP Large program \citep{Andrews..2021}. The ellipse in the bottom left illustrates the beam size of the observations.  \textbf{Bottom:} Azimuthal cut inside the gap at 48 au for the different models. It is shown that the undepleted case with a high accretion rate produces a noticeable bright structure in the disk. Moreover, there is some degeneracy between the local depletion in the CPD and accretion rate that hides the presence of the CPD at millimeter wavelengths, making it not noticeable}
    \label{fig:ALMA2}
\end{figure}

  In our synthetic predictions of the 1.3 millimeter dust continuum emission in Figure \ref{fig:ALMA}, under the right combination of parameters, the CPD will not show a significant difference in emission with respect to the surrounding material, i.e., the local dust depletion counteracts the viscous heating coming from the accretion. Only for the strong accretion scenario without dust depletion in the CPD, there is a strong local enhancement of millimeter emission at the planet's location. Nevertheless, as we show in the azimuthal cuts of Figure \ref{fig:ALMA2}, in the cases of low accretion and local dust depletion, the emission resembles an azimuthal absorption feature in millimeter dust continuum at the location of the planet. However, given the current observational capabilities, those changes can be distinguishable from the noise in the observations for deep integrations. With the exception of a high accretion rate, $\dot{M}>$10$^{-5}$ M$_{\rm Jup}$ yr$^{-1}$,  the emission enhancement are not above the 3\% level. On the contrary, when we assume a dust depletion in the CPD there is a significant decrement in the local emission to the levels of 10\%.

 The same exercise for the single-planet at 150 au under ideal conditions, i.e., without observational noises, would produce changes of $\sim$10\%; however, the disk is optically thin in millimeter wavelengths due to being embedded beyond the edge of the pebble disk,  having very low millimeter optical depth. Hence, even though the planet and accretion irradiation can produce significant relative changes, their absolute emission falls below the sensitivity threshold of the current state-of-the-art radio facility such as ALMA. In this case, no footprint or continuum emission feature is noticeable or detectable, although it may show an important effect in line emission.

\begin{figure}[h!]
    \centering
    \includegraphics[width=0.99\linewidth]{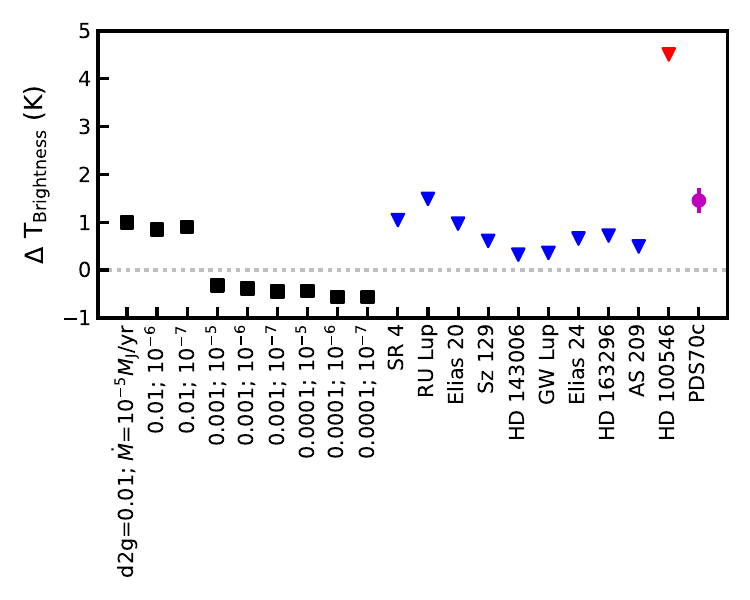}
    \caption{Expected brightness temperature deviations in our suite of models with respect to the baseline emission compared with deep high-resolution continuum data from ALMA programs. Black squares show that our models predict localized deviations with a brightness temperature lower than 2 K. Blue triangles show the 3$\sigma$ levels from  the \cite{Andrews..2021} limits, while the red triangle is the one from \cite{Perez..2020} and the red circle is the observed emission from PDS 70c.}
    \label{fig:Tbright}
\end{figure}

We compare the peak expected deviation in brightness temperature from our synthetic millimeter models to a planet at 50 au with the reported rms values in the literature for deep high-resolution dust continuum observations with ALMA in Figure \ref{fig:Tbright} \citep{Perez..2020, Benisty..2021, Andrews..2021}. The Figure shows that the expected deviations are marginally located at the 3$\sigma$ level detection in some cases even with very deep integrations. However, it is possible that some of the marginal proposed CPDs in the residual images of \cite{Andrews..2021} may be tracing different regimes in accretion and dust content within CPDs. 

In the case of PDS 70c, even though models in the literature predict the planet to be a Super Jupiter, the observed brightness temperature  shown in Figure \ref{fig:Tbright} is still consistent with the reported values by our predictions. From millimeter emission alone, it is hard to fully disentangle the accretion rate in the CPD of PDS 70c, but it is consistent with a dust-undepleted CPD. Nevertheless, multi-wavelength observations will help to constrain the accretion rate and the CPD content itself by breaking the degeneracies between those two variables.

 In the past, \cite{Wolf..Dangelo..2005} predicted a local bright emission considering a massive planet accreting at 5 au with current ALMA capabilities. However, in the bright sources observed with ALMA, such massive CPDs have not been observed. From current deep high-resolution observations, we can discard the presence of multiple high accretion rate scenarios without dust depletion because their local changes are within detectable sensitivity thresholds \citep{DSHARP..I}. Even though dust substructures can have multiple causes, under the assumption that a protoplanet is the cause of dust substructures, the low detection rate of CPDs with current facilities and surveys implies two possible scenarios: 
 \begin{enumerate}
     \item  The first explanation is that either high accretion rate states are temporary and relatively short, which can be explained by planets being formed in very short timescales. \cite{Lubow..Martin..2012} predict possible timescales if planetary growth undergoes accretion outbursts similar to FU Ori objects. They proposed that massive planets can undergo these states on timescales of 10$^3$-10$^5$ yr. As the detection of more protoplanets and characterizations of their masses and accretion rate become more ubiquitous, constraining the timescales of their formation will be more accurate, which will lead to a better understanding of planet formation.
     \item The second scenario is that planets accrete large grains in their CPDs on relatively short scales.  Significant depletion of large grains within translates into lower emission from optically thin CPDs, which would seem like azimuthally darker spots inside dust gaps due to the lack of effective millimeter emitters. Despite the fact there is one detection of a millimeter CPD from PDS 70c \citep{Benisty..2021}, it required deep integrations to observe a massive planet in a severely depleted disk with confirmed accretion tracers emission, so they are likely caught in their ongoing assembly. Regardless of PDS 70, in many other observations, deep campaigns for the detection of CPDs in dust continuum emission have been unsuccessful. Thus, millimeter emission from deep dust gaps, associated with massive planets, is already low, making the detection of azimuthal dips within them a hard task \citep{Andrews..2021}.
 \end{enumerate}

If we take a closer look at the protoplanet's orbital radius in the HD 163296 model by making an azimuthal cut at the separation of the 2 M$_{\rm Jup}$, we show that the planet creates a global emission effect at all azimuths, which follows from the global thermal increase. Nevertheless, there are still local modulations at the location of the planet depending on the accretion rate and local dust-to-gas ratio. As mentioned above, when there is a strong dust depletion within the Hill sphere of the planet, the local emission is lower, matching the one from the models without the planet present, which may look like an effective 'absorption' in the azimuthal profile. If we focus on the accretion rate, the viscous heating and shock emission only play a major role if the accretion rate is very high and the local dust-to-gas ratio is kept similar to the fiducial value of the disk. The lack of detections of CPDs in dust continuum images suggests that either planets do not have constant high accretion rates, i.e., they go through accretion outbursts analog to FU Ori objects; or there is a constant replenishment of dust grains to the CPD, keeping the dust-to-gas ratio roughly constant inside the CPD.

Doing the same exercise with the planet embedded beyond the pebble disk, where the thermal changes are more significant, the relative changes in emission become more important as well. However, the low local dust surface density causes the emission to fall below the detectable sensitivity thresholds of ALMA. Thus, even though there may be a significant temperature change beyond the pebble disk, it will not be noticeable by dust continuum ALMA observations. Nevertheless, such changes will manifest through line ratios and thermal changes in gas emission, which have the potential to be detectable under the right conditions. 

\section{Discussion}\label{sec:disc}

\subsection{Observational signatures of protoplanets and/or CPDs}

Former models explored the effect of depleted planet-carved gaps in protoplanetary disks and their observability \citep{Rosotti..2016}. Previous work in the literature \citep{Wolf..Dangelo..2005} showed that CPDs close to the star, if they were present, should be easier to detect in millimeter wavelengths with the best current observatories, which has not been the case except for PDS 70 b \citep{Isella..2019,Benisty..2021}. Further, \cite{Szulagyi..2018} tested the emission coming from planets at ALMA wavelengths, showing that massive planets have detectable footprints in the short wavelengths with ALMA. Such an approach was further expanded by \cite{Binkert..et..al..21} assuming a dust component through using a range of Stokes numbers. However, the physical conditions of material surrounding protoplanets can span over orders of magnitudes, broadening the range of Stokes numbers that a single particle size goes through while being accreted into the planet. \cite{Zhu..2018} mentioned that  CPDs may be detectable as long as they get replenished of large grains from the host disk. \cite{Andrews..2021} localized negative and positive residuals that can be consistent with the presence of CPDs; however, it is still hard to differentiate between real features and some artifacts not related to CPDs. Deeper resolved observations with higher sensitivity at different epochs should be able to confirm/discard the prevalence of the structures and their possible origin as CPDs.

Other methods of finding protoplanet candidates or CPDs through disk kinematics, line emission, or small grain content will be key in constraining the dust filtering and growth of planets in their early stages. It is expected that massive planets will produce velocity kinks in the channel maps of line emission \citep{Perez..2015, Perez..2018,  Pinte..2019}. The velocity kinks allow us to constrain the mass of a present protoplanet or to put at the very least an upper limit. The line emission asymmetries can help to constrain the dust depletion at the protoplanetary gap due to local chemical changes at these locations \citep{Cleeves..2015}.

Looking for CPD signatures beyond the pebble disk in millimeter dust continuum emission probes is hard given the low optical depth at millimeter wavelengths. Regardless of the lack of millimeter continuum emission, the models presented above show that there are significant thermal changes produced by the protoplanet emission, the CPD, and possible accretion onto the planet. Therefore, such changes may still be traceable by other means; temperature increases have the potential to lead to thermal desorption of otherwise absent volatiles, making them easier to observe due to the weak continuum emission and isolated emission. Moreover, the local density of dust can be low enough that the millimeter emission is negligible while making the CPD brighter than its surroundings in infrared wavelengths. A more thorough analysis of this behavior will be the topic of a subsequent paper.

Beyond the edge of pebble(millimeter) disks,  It is possible that the changes linked to the planet's emission may present themselves through the sublimation of key volatiles \citep{Cleeves..2015}. More sensitivity and deeper studies of line emission in their optically thin regions may point to the presence of, otherwise hidden, circumplanetary disks. Shocked gas during the accretion of massive planets can reach very hot temperatures ($>$3000 K) producing optically thick atomic blankets hiding the protoplanet emission \citep{Szulagyi..mordasini..2017..shocks}. Shocks can also play a major role in local photochemistry through photodissociation and photodesorption as well, which could be probed by the emission of other tracers such as SO, or CI \citep{Alarcon..2022, Law..2023}.   A tentative detection of a CPD beyond the pebble disk has been suggested in the AS 209 disk, where there is a local enhancement of the predicted gas density, but also a slight increase in the temperature of the emission measured by CO isotopologues \citep{MAPS..XXI..Bae}. The measurement of this thermal increase is a key ingredient in determining the CPD mass, and therefore, the properties of the protoplanet.

The observable effects in the millimeter dust continuum emission of the Jovian planet at 150 au are negligible. However, in most cases, when there is a lack of significant replenishment and/or a strong accretion rate, there will be a significant gas temperature increase which should have a strong imprint on the local gas emission. So, even though the dust millimeter continuum emission will not show strong observable footprints, the footprints of a planet should be traceable by analysis of molecular line emission of different transitions for the same tracers as well.

Overall, when high accretion rates and strong dust depletion are combined, important effects are expected in both physics and chemistry. Those changes include multiple conditions such as high energy radiation fields, local temperature, thermal desorption rates, line ratios, line emission, and reaction rates of the gas chemistry.

\subsection{The low number of CPD detections}

The low number of CPDs detected to date through millimeter continuum observations can point to a lack of millimeter grains if massive planets are actually carving the dust gaps or low accretion rates. A possible reason behind the lack of CPD detections is that massive planets accrete the solid reservoir in CPDs in very short timescales, together with some level of dust replenishment or very efficient dust growth. Nevertheless, the net effect of the blackbody emission on their surroundings, in particular in dust-depleted gaps, may be leading to an overestimation of the local dust surface density. This means that the actual local depletion may be stronger, which can have strong implications for our understanding of the values of disk viscosity, disk shape, and predicted planetary masses from observations.

Another possible explanation for the lack of CPD detections is that most of the observed dust substructures are not actually being carved by massive planets, but actually being caused by other mechanisms and/or physical instabilities (see \cite{Bae..2023} for a detailed overview of substructure-forming mechanisms). Many of these instabilities will lead to structures resembling the ones produced by massive planets. i.e., gaps, rings, and spirals. It is only through deep observation and/or multiple analytical approaches involving continuum and line emission coupled with modeling that the presence of a planet, or the absence of instabilities, can be tackled.

\subsection{Hydrodynamics Simulation Caveats}

In our initial hydrodynamics simulations we assumed isothermal structures, it is important to note that the actual distribution of infalling material into the planet will depend on whether an adiabatic or isothermal equation of state is chosen, producing a more spherical or flat matter distribution respectively \citep{Szulagyi..2016,Fung..2019, Paardekooper}. The same changes in the assumed thermodynamics caused different densities around the planet. The isothermal solution produced lower densities in the CPD than the adiabatic regime. Thus, depending on the thermodynamical behavior of the gas around the protoplanet the signature may change due to optical depth effects.

We also opted not to use \texttt{FARGO3D} multifluid module to study the dust distribution, in particular, because we focus on the regions close to the protoplanet. Multifluid modules generally rely on the Stokes number, St, although the Stokes number depends on the stopping time and Keplerian velocity and it does not trace a uniform grain size. Since the stopping time depends on the local surface density and the Keplerian velocity, the grain size can range more than one order of magnitude for the same Stokes number. Thus, it is not a reliable tracer of the dust content in the CPD, unless a dedicated study is conducted, which goes beyond the scope of this paper. \cite{Krapp..2022} and \cite{Krapp..2024a} emphasize the need for global radiation hydrodynamics to fully understand the dust distribution and dynamics in the vicinity of the planet. \cite{Krapp..2024b} also presented a new solver for particle-size distribution studies in multifluid modules to study dust dynamics.

\section{Summary}\label{sec:summary}

We present the thermal changes produced by including the emission of an accreting planet and a simple prescription for its associated shocks on its host protoplanetary disks for cases when the planet is embedded inside the optically thick pebble disk, and one when the planet is located in the more diffuse less dense outer disk.

From post-processing 3D hydrodynamics with radiative transfer, we show that the radiation of a massive planet increases further the local and global temperature beyond the effect of the dust substructures it produces. For low values of accretion rate, the changes in local millimeter emission are not noticeable. Moreover, we found that a high accretion rate combined with a high depletion of large grains can hide the emission of a hotter CPD. Hence, CPDs brightness temperature may look hotter/colder than its surroundings depending on the local level of dust depletion. 

The temperature differences become more important beyond the millimeter disk as the stellar viscous heating becomes negligible and the disk gets cooler. However, the dust surface density is low enough that the emission is optically thin. Hence, the millimeter emission, and its changes, fall below the sensitivity threshold of current state-of-the-art facilities such as ALMA. Nevertheless, the thermal effects can potentially be detectable in gas emission or temperature tracers from protoplanetary disks.

\begin{acknowledgments}
F.A, E.A.B. acknowledge support from NSF AAG Grant \#1907653. 
\end{acknowledgments}

%


\software{astropy \citep{AstropyI,AstropyII,AstropyIII}, \texttt{FARGO3D} \citep{FARGO3D},
\texttt{RADMC-3D} \citep{RADMC-3D},
numpy \citep{numpy},
matplotlib \citep{matplotlib},
cmasher \citep{cmasher}
          }



\appendix

\section{Appendix information}


\bibliography{ref}{}
\bibliographystyle{aasjournal}



\end{document}